\documentclass[letterpaper,10pt, twocolumn]{article}
\usepackage{amsmath}
\usepackage{latexsym,amssymb}
\usepackage{epsfig}
\usepackage{graphicx}
\usepackage{bm}
\usepackage{setspace}

\newcommand{\psistate}{\left|\Psi\right>}

\begin{document}
\markright{Theory Canada IV}

\title{{\large{\bf On the black hole singularity issue in loop quantum gravity}}}
\author{\small A. DeBenedictis$^{*}$
\\ \it{\small Pacific Institute for the Mathematical Sciences, Simon Fraser University Site} \\
{\small and} \\ \it{\small Department of Physics,} \\
\it{\small Simon Fraser University, Burnaby, British Columbia, V5A
1S6, Canada}}
\date{{\small May 28 2008}}

\twocolumn[
  \begin{@twocolumnfalse}
    \maketitle
\vspace{-7.0mm}
    \begin{abstract}
 \noindent This paper presents a brief overview on the issue of singularity resolution in loop quantum gravity presented at the Theory Canada IV conference at the Centre de Recherches Math\'{e}matiques at the Universit\'{e} de Montr\'{e}al (June 4-7, 2008). The intended audience is  theoretical physicists who are non-specialist in the field and therefore much of the technical detail is omitted here. Instead, a brief review of loop quantum gravity is presented, followed by a survey of previous and current work on results concerning the resolution of the classical black hole singularity within loop quantum gravity. \\
\rule{\linewidth}{0.20mm} \\
\vspace{-3.0mm}
\begin{center}
 {\bf R\'{e}sum\'{e}}
\end{center}
\noindent Nous pr\'{e}sentons une br\`{e}ve revue portant sur la r\'{e}solution du probl\`{e}me de
singularit\'{e} en gravit\'{e} quantique \`{a} boucles, destin\'{e}e \`{a} la conf\'{e}rence Th\'{e}orie Canada 4
tenue au Centre de Recherches Math\'{e}matiques de l'Universit\'{e} de Montr\'{e}al (4-7 juin 2008). L'audience
vis\'{e}e est celle de physiciens th\'{e}oriques qui ne sont pas sp\'{e}cialistes du domaine et par
cons\'{e}quent, nous omettons la plupart des d\'{e}tails techniques. \`{A} la place, nous présentons une
br\`{e}ve revue de la gravit\'{e} quantique \`{a} boucles, suivie d’un survol des r\'{e}sultats ant\'{e}rieurs
et pr\'{e}sents concernant la r\'{e}solution de la singularit\'{e} classique du trou noir dans le cadre de
la gravit\'{e} quantique \`{a} boucles. \\
\rule{\linewidth}{0.5mm}
{\footnotesize * adebened@sfu.ca}
    \end{abstract}
  \end{@twocolumnfalse}
  ]

\section{{\normalsize Introduction}}
\vspace{-2.0mm}
General relativity is arguably the most successful theory of gravity to date. Even after more than 90 years, the theory is extremely robust at predicting solar system scale physics to astounding accuracy. On much larger scales, general relativity supplemented by various exotic sources, such as dark matter or dark energy, seems to explain very well the properties of the universe we live in. However, on the extremely small scale, or equivalently very high energy scales, there is precious little we know about gravity.
In this sector we have to heavily rely on what we \emph{expect} gravity to behave like in this regime. At very high energies we expect that classical general relativity will get modified in some way by quantum effects. However, the quantization of the gravitational field is not at all straight-forward, as full diffeomorphism invariance requires a different approach than those applied to conventional field theories. Other problems associated with the quantization of gravity may be found in \cite{ref:structural}, which also presents a nice survey of the field in general.

One potential candidate for a theory of quantum gravity is known as loop quantum gravity which, over the past couple of decades, has become a strong contender in this arena. The loop quantum gravity program incorporates diffeomorphism invariance as demanded by general relativity and, as will be outlined below, is in fact a quantization of general relativity. An excellent (but technical) overview of loop quantum gravity may be found in \cite{ref:Rlivrev}. Essentially, what arises out of loop quantum gravity is a theory of quantum geometry, which is quite appealing from the point of view of general relativity. 

Having a candidate theory at hand, the natural question to ask is what theoretical (in the absence of experimental) issues do we wish the theory to resolve? There are probably many but two that come to mind are the explanation of black hole entropy and the elimination of the classical singularities such as those encountered in cosmology and black hole physics. 

Regarding the entropy, there is still debate on the actual source of this entropy. One
belief is that the source is strictly gravitational in origin. That is, one should be able to define microstates in a full
quantum theory of gravity which, when counted, should yield the correct entropy law. It has been shown that for spherical black holes the correct entropy law is derived within loop quantum gravity \cite{ref:entrev1} - \cite{ref:entrev6}. The result has been extended to higher genus black holes, where it was found that the sub-leading term to the entropy is genus dependent \cite{ref:ourent}. This is in agreement with studies of black hole entropy of higher genus black holes that have utilised non-quantum gravity techniques 
\cite{ref:vanzo}, \cite{ref:mans} \cite{ref:liko}. Progress from the spin-foam approach has also been made \cite{ref:manuel1}, \cite{ref:manuel2}.

On the singularity issue, most of the work in loop quantum gravity has been done within the paradigm of loop quantum cosmology (LQC). In LQC all but the classically relevant degrees of freedom are frozen out first and then the resulting, symmetry reduced, system is quantized. There are, of course, some limitations inherent in the various approaches to LQC (for example, see \cite{ref:brunthiem}). However, studies to date indicate that the quantum corrections become important as one approaches the big bang singularity. The singularity itself is therefore replaced with a smooth bounce. 

There are many studies in the literature on loop quantum cosmology. We do not survey the vast literature here due to space limitations. However, the interested reader is referred to \cite{ref:LQCLivrev} and the references therein. Instead, here we chose to briefly discuss the progress on the black hole front, which utilises many of the same techniques used in LQC. 

It should be stressed that the intended audience is not the expert, the Theory Canada conferences being made up of theoretical physicists from a variety of fields and backgrounds. Instead, only the main points are presented, with little technical detail. However, it is hoped that this will give the reader an overview of loop quantum gravity as well as the flavor of the various approaches to the problem of singularities within this theory. Of current research interest to us is the extension of these studies to various other singular solutions of general relativity. Progress on this front has been made in \cite{ref:BKDsing} where singularities have been shown to be avoided in black holes other than spherical.

\section{{\normalsize A brief introduction to loop quantum gravity}}
We begin by considering the Einstein-Hilbert action of general relativity:
\begin{equation}
I\left[g_{\mu\nu}\right]= \frac{1}{8\pi} \int\,d^{4}x \, \sqrt{-g}\,R(g_{\mu\nu}) + I_{M}(g_{\mu\nu})\,, \label{eq:EHact}
\end{equation}
where $g$ is the determinant of the metric tensor, $g_{\mu\nu}$, and $R(g_{\mu\nu})$ is the Ricci scalar, which is to be viewed as a functional of the configuration variable, which is the metric tensor. The quantity $I_{M}(g_{\mu\nu})$ represents the matter field action. Demanding that the variation of this action with respect to $g_{\mu\nu}$ vanishes gives rise to the Einstein field equations,
\begin{equation}
 R_{\mu\nu}-\frac{1}{2}R\,g_{\mu\nu}=8\pi T_{\mu\nu}\,, \label{eq:einsteq}
\end{equation}
with $R_{\mu\nu}$ being the Ricci tensor and $T_{\mu\nu}$ the stress-energy tensor of the matter field. 

\begin{figure}[!ht]
\begin{center}
\includegraphics[bb=0 0 910 354, clip, scale=0.28, keepaspectratio=true]{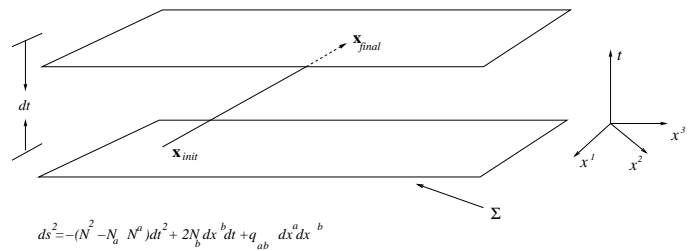}
\caption{{\small ADM decomposition of space-time into ``space'' and
``time''. In this decomposition, one studies how the three geometry of $\Sigma$ evolves with time $t$.}} \label{fig:admdecomp}
\end{center}
\end{figure}

We will consider here the canonical quantization of the Einstein-Hilbert action within the paradigm of loop quantum gravity. A covariant form also exists, giving rise to spin-foam models, so named due to the resemblance of the Feynman diagram analogs to a foam of bubbles. However, for an introduction the canonical approach is much more appropriate. In the canonical approach one splits space-time into ``space'' and ``time'' by writing the line element associated with the metric as:
\begin{align}
 ds^{2}:=g_{\mu\nu}dx^{\mu}dx^{\nu}=&-\left(N^{2}-N_{a}N^{a}\right)dt^{2} \nonumber \\
 &+2N_{b} dx^{b} dt +q_{ab}dx^{a}dx^{b}\,. \label{eq:admmet}
\end{align}
Here, $N$ and $N_{b}$  are the lapse function and shift vectors respectively and $q_{ab}$ is the metric on the three-dimensional hyper-surfaces, $\Sigma$. These quantities are shown schematically in figure \ref{fig:admdecomp}.

In the variables of equation (\ref{eq:admmet}), the action (\ref{eq:EHact}) may be written as
\begin{eqnarray}
 I&=&\frac{1}{16\pi} \int dt \int_{\Sigma}dx^{3} \left[\Pi^{ab}\dot{q}_{ab} +2N_{b} \nabla^{(3)}_{a} \left(q^{-1/2} \Pi^{ab}\right) \right.\nonumber \\
 &&\left. +Nq^{1/2}\left(R^{(3)} -q^{-1}\Pi_{cd}\Pi^{cd} +\frac{1}{2} q^{-1}\Pi^{2}\right)\right], \label{eq:ADMact}
\end{eqnarray}
with $\Pi:=\Pi^{ab}q_{ab}$, $\Pi^{ab}:=q^{-1/2}\left[K^{ab}-K\,q^{ab}\right]$, $K:=K^{ab}q_{ab}$. $K_{ab}$ is the extrinsic curvature tensor, $R^{(3)}$ the Ricci scalar of the three-dimensional hyper-surface and $q$ the determinant of $q_{ab}$ with over-dots denoting differentiation with respect to the slicing time, $t$. In this scheme, the field variable is $q^{ab}$ and its conjugate momentum is $\Pi^{ab}$. The resulting equations of motion are:
\begin{eqnarray}
  2\nabla^{(3)}_{a}\left[q^{-1/2}\Pi^{ab}\right]=0 =:-V^{b}\:, \nonumber \\
  q^{1/2}\left[R^{(3)}-q^{-1} \Pi_{ab}\Pi^{ab} + \frac{1}{2}q^{-1}\Pi^{2}\right]=0=:-S\:. \nonumber
 \end{eqnarray}
The action in (\ref{eq:ADMact}) leads to a Hamiltonian density:
 \begin{eqnarray}
  \mathcal{H}_{G}=\frac{1}{16\pi}\Pi^{ab}\dot{q}_{ab} -\mathcal{L}_{G} =N_{b}V^{b}+N\:S\;, \nonumber
 \end{eqnarray}
 with the following symplectic structure:
 \begin{eqnarray}
\left\{\Pi^{ab}(\mathbf{x}),\, q_{cd}(\mathbf{y})\right\}=16\pi \delta^{a}_{\,(c}\delta^{b}_{\;d)} \delta(\mathbf{x},\,\mathbf{y}), \nonumber \\
  \left\{\Pi^{ab}(\mathbf{x}),\,\Pi^{cd}(\mathbf{y})\right\} = 0 = \left\{q_{ab}(\mathbf{x}),\, q_{cd}(\mathbf{y})\right\}\:. \nonumber 
 \end{eqnarray}
 After quantization, this leads to the well-known Wheeler-DeWitt equation:
 \begin{eqnarray}
\left\{\left[-q_{ab}q_{cd}+\frac{1}{2}q_{ac}q_{bd}\right]\frac{\delta}{\delta q_{ac}} \frac{\delta}{\delta q_{bd}} +q\,R^{(3)}\right\} \Psi(q)=0\,. \label{eq:WdW} 
 \end{eqnarray}
The above equation was first derived by DeWitt \cite{ref:WdWdewitt} and Wheeler \cite{ref:WdWwheeler} in the 1960s. There are some problems associated with the approach leading to (\ref{eq:WdW}). Some of the problems are associated with all canonical approaches to quantum gravity, such as the ``issue of time'', which leaves open exactly how to determine evolutions in a theory where time is not manifest in the quantum equations. Other issues with the Wheeler-deWitt formalism involve inconsistencies in some transition amplitudes in a path integral formulation of the theory. Also, the three-metric configuration variable does not appear as a gauge field in the theory, which means many of the techniques used in gauge field theories are not applicable here.  As well, as will be briefly mentioned below, the singularities present in classical gravitation are not eliminated within the WdW formalism.

In order to make contact with gauge theories, Ashtekar formulated the gravitational action in terms of \emph{densitized triads} $E_{\;i}^{a}$ \cite{ref:ashtriad}, which are related to the 3-metric components via
\begin{equation}
   q\,q^{ab}=E_{\;i}^{a}E_{j}^{b}\delta^{ij}\,, \nonumber
 \end{equation}
and are related to the standard triad, $e^{j}_{\;b}$, by
\begin{equation}
 E^{a}_{\;i}=\frac{1}{2}\epsilon^{abc}\epsilon_{ijk} e^{j}_{\;b}e^{k}_{\;c}\,. \nonumber
\end{equation}

In terms of these new variables, the $3+1$ Einstein-Hilbert action may be cast in the form:
\begin{align}
 \nonumber I=& \frac{1}{8\pi} \int dt\int_{\Sigma} \left[E^{a}_{\;i} \dot{K}^{i}_{\;a} -N_{b}V^{b} -NS \right. \\
& \left. \hspace{2.0cm} -N^{i} \epsilon_{ijk}E^{aj}K^{k}_{\;a} \right]d^{3}x\;, \nonumber \\
\nonumber K^{i}_{\;a}:=& \frac{1}{\sqrt{\det (E)}} K_{ab} E^{b}_{\;j} \delta^{ij} \: .
 \end{align}
The canonically conjugate variables are now $E^{a}_{\;i}$ and ${K}^{i}_{\;a}$.
The transformation group of the triad in $\Sigma$ is $SO(3)$ and as such, so is that of its conjugate momentum, $K^{i}_{\;a}$. The natural $SO(3)$ connection compatible with the vanishing of the covariant derivative of the triad is given by $\Gamma^{i}_{\;a}$ as in
\begin{equation}
 \partial_{[a}e^{i}_{\;b]}+\epsilon^{i}_{\;jk}\Gamma^{j}_{[a}e^{k}_{\;b]}=0\,.
\end{equation}

We can also define another $so(3)$ connection $A^{i}_{\;a}$ by the linear combination:
 \begin{eqnarray}
  A^{i}_{\;a}:=\gamma K^{i}_{\;a}+ \Gamma^{i}_{\;a}\;, \nonumber
 \end{eqnarray}
 where $\gamma$ is known as the \emph{Immirzi parameter}, which must be determined in some way and is therefore a type of quantization ambiguity at this point. The new connection $A^{i}_{\;a}$ turns out to also be conjugate to $E^{i}_{\;a}$ and this pair, $\left(A^{i}_{\;a}\,,\,E^{i}_{\;a}\right)$, make up the Ashtekar-Barbero variables on which loop quantum gravity is based. In terms of these new variables, the action is written as
 \begin{align}
 I=&\frac{1}{8\pi} \int dt\int_{\Sigma} \left[E^{a}_{\;i} \dot{A}^{i}_{\;a} -N_{b}V^{b} -NS  -N^{i} G_{i} \right]d^{3}x\;, \label{eq:Aact}\\
  G_{i}:=&\partial_{a}E^{a}_{\;i} +\epsilon_{ij}^{\;\;\;k}A^{j}_{\;a}E^{a}_{\;k}\:. \nonumber 
\end{align}
Here, 
\begin{subequations}
\begin{align}
V_{b}:=&E^{a}_{\;j}F_{ab}^{j} -(1+\gamma^{2})K^{i}_{\;b}G_{i}\,, \\
S:=&\left[\mbox{det}(E)\right]^{-1/2}E^{a}_{\;i}E^{b}_{\;j}\left[\epsilon^{ij}_{\;k}F^{k}_{\;ab}-2\left(1+\gamma^{2}\right)K^{i}_{\;[a}K^{j}_{\;b]}\right]\,,
\end{align}
\end{subequations}
with $F^{i}_{\;ab}:= \partial_{a}A^{i}_{\;b} -\partial_{b} A^{i}_{\;a} +\epsilon^{i}_{\;\;jk}A^{j}_{\;a}A^{k}_{\;b}\,$.

The corresponding symplectic structure is given by:
 \begin{eqnarray}
 \left\{E^{a}_{\;j}(\mathbf{x}),\,A^{i}_{\;b}(\mathbf{y})\right\}=8\pi \gamma \delta^{a}_{\;b} \delta^{i}_{\;j}\,\delta(\mathbf{x}, \mathbf{y})\,, \nonumber \\
 \left\{E^{a}_{\;i}(\mathbf{x}),\, E^{b}_{\;j}(\mathbf{y})\right\}=\left\{A^{j}_{\;a}(\mathbf{x}),\,A^{i}_{\;b}(\mathbf{y})\right\}=0\,. \nonumber
 \end{eqnarray}
To the above system one could also add matter couplings by supplementing the action with a matter term and quantizing appropriately. The problems associated with (\ref{eq:WdW}) are not present in this representation and it can be seen that $A^{i}_{\;b}$ enters like a gauge field.

Variation of the action with respect to the shift-vector, $N_{b}$, and lapse function, $N$, supplemented with the fixing of rotational freedom leads to what are known as the vector, scalar and Gauss constraints respectively:
\begin{subequations}
 \begin{align}
  V_{b}=&0\,, \\
S=&0\,, \\
G_{i}=&0\,
 \end{align}
\end{subequations}
(the explicit forms are given below.) Note that now there is a constraint equation for $G_{i}$ (the Gauss constraint). This reflects the fact the triad possesses a rotational freedom; one can choose different frames locally by rotating the triad. This redundancy is eliminated in the new Gauss constraint.

It is customary to work in $SU(2)$, the universal covering group, since the Lie algebra is the same and this change does not affect the constraints. It is also useful from the point of view of bridging quantum gravity with standard gauge field theory. Therefore the connection $A$ is now to be viewed as an $SU(2)$ connection.

The constraint equations that arise, plus the rotational degree of freedom of the triad give rise to the following equations after quantization:
\begin{subequations}
\begin{align}
 \hat{G}_{i}\psistate=&\widehat{D_{a}E^{a}_{\;i}}\left|\Psi\right> =0,  \\
 \hat{V}_{b}\psistate =& \left[\widehat{E^{a}_{\;i}F^{i}_{\;ab}} -(1+\gamma^{2}) \widehat{K^{i}_{\;b}G_{i}}\right]\psistate=0,  \\
\hat{S}\psistate=&\frac{1}{\sqrt{\det(\hat{E})}} \widehat{E}^{a}_{\;i}\widehat{E}^{b}_{\;j} \nonumber \\
&\times \left[\epsilon^{ij}_{\;\;\;k}\widehat{F}^{k}_{\;ab} -2(1+\gamma^{2}) \widehat{K}^{i}_{\;[a}\widehat{K}^{j}_{\;b]} \right]\psistate=0\,,   
\end{align}
\end{subequations}
where $D_{a}$ is the gauge covariant derivative given in the second line of equation (\ref{eq:Aact}).

One can construct functionals $\Psi(A)$ which are functionals of $A$ and are square-integrable with respect to the Ashtekar-Lewandowski-Baez measure \footnote{Strictly speaking it is the \emph{holonomies} of $A$ which are of fundamental importance. However, as we are omitting technical details, this can be ignored at this stage. It is important to note however that with the concept of holonomy, paths become an important feature of the theory.}:
  \begin{eqnarray}
   \left<\Psi,\,\Phi\right>=\int d\mu[A]\, \overline{\Psi}[A]\, \Phi[A]\,. \nonumber 
  \end{eqnarray}
This measure is gauge as well as diffeomorphism invariant. The corresponding Hilbert space possesses some peculiar properties which we do not address here. The interested reader may refer to \cite{ref:Rlivrev} and references therein. 

Before continuing, we summarize a few points about loop quantum gravity that the reader may be curious about (recalling that the intended audience is a theorist who does not specialize in loop quantum gravity):
\begin{enumerate}
  \item The scheme is background independent and respects diffeomorphism invariance. The choice of time slicing is arbitrary and does not affect the physics.   
  \item A superpartner can be accommodated and therefore supersymmetry can be incorporated. This has been done \cite{ref:superLQG}.
  \item Instead of the Einstein-Hilbert action, one can accommodate geometric actions made up of arbitrary curvature invariants. The scheme is generally similar to that outlined above.
  \item Higher dimensions can be accommodated.
\item The theory is divergence free.
 \end{enumerate}
It should be noted that 2, 3, and 4 are \emph{not required} but simply can be accommodated.

\subsection{{\small Geometric operators}}
\begin{figure}[h!t]
\begin{center}
\includegraphics[bb=0 0 612 414, clip, scale=0.40, keepaspectratio=true]{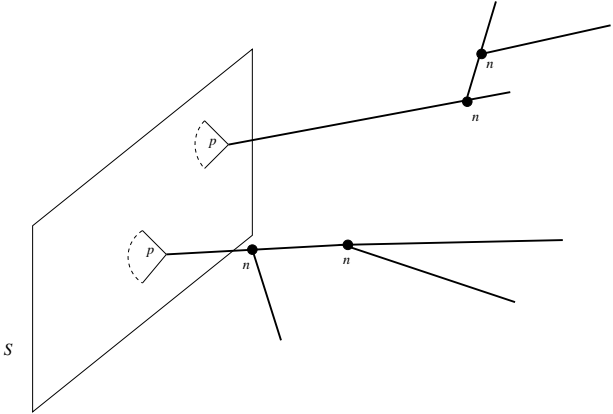} 
\caption{{\small A spin-network ``puncturing'' a surface. Each part of the spin-network puncturing the surface endows it with a certain area and local angular deficit, yielding non-trivial geometry for the surface. The nodes ($n$) are associated with volumes.}} \label{fig:2}
\end{center}
\end{figure}

Having discussed (very briefly) the loop quantum gravity program we now turn our attention to important operators within the theory. As this is a theory of quantum geometry, areas and volumes are of particular importance. ``Classical'' areas ($Ar$) and volumes ($V$) may be constructed from the triad and, equivalently, from the densitized triad:
\begin{align}
 Ar=&\int_{\mathcal{S}}\,d^{2}\sigma\, \sqrt{E^{a}_{\;i}E_{b}^{\;i}n_{a}n^{b}} \,, \nonumber \\
V=&\int_{M}\,d^{3}x\, \left[\left|\epsilon_{abc}\epsilon^{ijk} E^{a}_{\;i} E^{b}_{\;j} E^{c}_{\;k}\right| \right] \,. \nonumber
\end{align}
Since these involve $E$, they become operators in the quantum theory.
The area operator possesses the following eigenvalues:
\begin{equation}
\widehat{(Ar)}\psistate=8\pi l_{p}^{2}\gamma\sqrt{j(j+1)}\psistate \,, \label{eq:areaspectrum}
\end{equation}
$l_{p}$ being the Planck length. The interpretation of the above equation is as follows; The state functions of the theory describe a \emph{spin-network}. This arises from the fact that the relevant operator made up of the configuration variable is not the connection $A$ itself, but the holonomy of $A$ along an oriented path $e$. This holonomy is a functional of the connection $A$ and allows for the parallel transport of spinors along $e$. Combinations of these paths make up the spin-network and each edge ($e$) of the path carries with it a half-integer spin, denoted by the symbol $j$ in (\ref{eq:areaspectrum}). When a segment of the spin network penetrates a surface at a point $p$, it endows the surface with an amount of area related to its spin. The specific amount is given by the eigenvalue in (\ref{eq:areaspectrum}). The total area of the surface is given by adding all of the eigenvalues of the penetrating network segments over the surface $S$. A simple diagram illustrating this is shown in figure \ref{fig:2} \footnote{We are making an assumption here regarding how the spin-network pierces the surface $\mathcal{S}$. The general case yields eigenvalues which are slightly more complicated than (\ref{eq:areaspectrum}).}.

The very interesting issue is that this theory predicts that, on the small scale, space is \emph{discrete}! It should be emphasized that this is a \emph{prediction} of the theory and is not put in ``by hand''. The volume operator is more complicated to deal with but also predicts that volumes are quantized. (Volumes are associated with the nodes of the spin-network.)

\section{{\normalsize Singularity resolution}}
It has long been thought that any viable theory of quantum gravity should eliminate the singularities predicted by the classical theory. Specifically, the space-like singularities of black hole physics and standard cosmology should be removed. If a theory is successful in this endeavor, a natural question to ask is what replaces the singularity? We will briefly outline progress and results within loop quantum gravity on this front here.

Given the complexity of the situation, studies are generally carried out in \emph{symmetry reduced} models. That is, one first imposes certain symmetries at the classical level and then quantizes the symmetry reduced model in the hope that the essential features of the full theory are retained.

There are several approaches to studying black hole singularities in loop quantum gravity. We will first be concentrating on the approach used in \cite{ref:AandB} (related works may be found in \cite{ref:lqgsing1} and \cite{ref:lqgsing2} and references therein.)

To date, the Schwarzschild black hole has received by far the most attention. The first step is the symmetry reduction to a symmetric system appropriate for the black hole interior $(R\times S^{2})$. The connection and triad respecting this symmetry are given by
\begin{subequations}
\begin{align}
 A=&A_{R}\tau_{3}\,dR +\left(A_{1} \tau_{1} +A_{2}\tau_{2}\right)d\theta +\left[\left(A_{1}\tau_{2}-A_{2}\tau_{1}\right)\sin\theta \right. \nonumber \\
& \left. +\tau_{3}\cos\theta\right]d\varphi\, , \\
E=& E^{R}\tau_{3}\sin\theta\, \frac{\partial}{\partial R} + \left(E^{1} \tau_{1} + E^{2} \tau_{2}\right)\sin\theta\frac{\partial}{\partial\theta} \nonumber \\
&+\left(E^{1}\tau_{2} -E^{2}\tau_{1}\right) \frac{\partial}{\partial\varphi}\, , 
\end{align}
\end{subequations}
where the $\tau_{i}$ denote the standard $su(2)$ basis. (Note that the indices $1$ and $2$ on the quantities $A$ and $E$ are not directly related to the $1$ and $2$ indices on the $\tau$ matrices.) In this reduced scenario, the vector constraint is automatically satisfied and
the Gauss contraint is satisfied if $A_{1}E^{2}-A_{2}E^{1}=0$. In the sequel $A_{1}$ and $E^{1}$ are set equal to zero to satisfy this constraint. The Hamiltonian constraint, which is a combination of the Gauss, vector and scalar constraints yields:
\begin{equation}
 \mathcal{H}=-\frac{N}{8\pi \gamma^{2}} \frac{\mbox{sgn}E^{R}}{\sqrt{|E^{R}|}E^{2}} \left[\left(A_{2}^{2} + \gamma^{2}\right)(E^{2})^{2} +2 A_{R}E^{R}A_{2}E^{2}\right]\,. \label{eq:frozenham}
\end{equation}
It can be shown (still dealing at the classical level), that the following quantities solve the equations of motion \cite{ref:AandB}:
\begin{eqnarray}
A_{2}=&\pm\gamma\sqrt{(2M-T)/T}\,,&E^{2}=E^{2}_{0}\sqrt{T(2M-T)} \nonumber \\
A_{R}=&\mp A_{R}^{0}\gamma \frac{M}{T^{2}}\,, & E^{R}=\pm T^{2}\,. \nonumber
\end{eqnarray}
(A coordinate re-scaling has tacitly been assumed.)

It turns out (not surprisingly) that the above (densitized) triad is compatible with this line element's metric:
\begin{equation}
 d\sigma^{2}=\left(\frac{2M}{T}-1\right)dR^{2} + T^{2}\,d\theta^{2} +T^{2}\sin^{2}\theta\,d\varphi^{2}\,, \nonumber
\end{equation}
which is the $T$-domain, or black hole interior, of the Schwarzschild space-time.

Next, the symmetry reduced system is quantized via:
\begin{equation}
 \hat{E}^{2}:=-i\gamma l_{p}^{2} \frac{\partial}{\partial A_{2}}\;, \;\;\;\;\; \hat{E}^{R}:=-2i\gamma l_{p}^{2} \frac{\partial}{\partial A_{R}} \; , \nonumber
\end{equation}
with the following spectrum:
\begin{equation}
\hat{E}^{2} \left|\mu,\,\tau\right>= \frac{1}{2}\gamma l_{p}^{2} \mu \left|\mu,\,\tau\right>,\;\;\;\;  \hat{E}^{R} \left|\mu,\,\tau\right>=\gamma l_{p}^{2} \tau \left|\mu,\,\tau\right>\,. \nonumber
\end{equation}
Here, $\tau$ and $\mu$ represent the coordinate length of the holonomy curve in the $R$ and angular directions respectively in the state vectors.

The volume operator is also constructed:
\begin{subequations}
\begin{align}
&\hat{V}=4\pi|\hat{E}^{2}| \sqrt{|\hat{E}^{R}|}\;, \\
& \mbox{with eigenvalues} \;\;\; \;v=2\pi\gamma^{3/2}l_{p}^{3} |\mu| \sqrt{|\tau|}\,.
\end{align}
\end{subequations}
Also, in what follows, a general state will be expanded in eigenfunctions of the triad as:
\begin{equation}
 \psistate = \sum_{\mu,\tau} C_{\mu\tau} \left|\mu,\,\tau\right>\,, \nonumber
\end{equation}
so that it is sufficient to study effects on $\left|\mu,\,\tau\right>$ and on the expansion coefficients $C_{\mu\tau}$. 

Before proceeding we should note that there are several ways to indicate potential singularity avoidance. One is to study the quantum analogues of the curvature. Another is to study the evolution equations and see if they are smooth (non-singular) where the classical singularity exists ($T=0$). Both were analyzed in \cite{ref:AandB}, whose results we quote here. For the curvature argument, we require the co-triad, $\omega$:
\begin{equation}
 \omega=\omega_{R}\tau_{3}\,dR + \omega_{2} \tau_{2}\,d\theta -\omega_{2}\tau_{1}\sin\theta\,d\varphi\,. \nonumber
\end{equation}
Classically, the co-triad involves inverse powers of the triad and is useful for forming quantities measuring curvature and in the Hamiltonian constraint, which also contains inverse powers of the triad. Explicitly, the classical co-triad is given by:
\begin{equation}
\omega_{R}=\mbox{sgn}\left(E^{R}\right) \frac{|E^{2}|}{\sqrt{|E^{R}|}}\;,\;\;\;\omega_{2}=\mbox{sgn}\left(E^{2}\right) \sqrt{|E^{R}|}\, . \label{eq:cotr}
\end{equation}
To calculate the quantum quantities appearing in the corresponding operator, $\hat{\omega}_{R}$, which is the component involving the inverse power of the triad, one first rewrites the classical co-triad in a classically equivalent manner and creates the corresponding quantum operator by analogy \cite{ref:thiemtrick}:
\begin{align}
 \omega_{R}=&\frac{1}{2\pi\gamma}\mbox{Tr}\left\{\tau_{3} h_{R}\left[h_{R}^{-1},\, V\right]\right\} \nonumber \\
& \Longrightarrow \hat{\omega}_{R}=-\frac{i}{2\pi\gamma l_{p}^{2}}\mbox{Tr}\left\{\tau_{3} \hat{h}_{R}\left[\hat{h}_{R}^{-1},\, \hat{V}\right]\right\}\,.
\end{align}
Here $\hat{h}_{R}$ is the holonomy (operator) of $A_{R}$, which has a well defined inverse operator.
With this definition it can be shown that
\begin{eqnarray}
\hat{\omega}_{R} \left|\mu,\,\tau\right> &=&\frac{1}{4\pi\gamma l_{p}^{2}} \left(v_{\mu,\,\tau+1} - v_{\mu,\,\tau-1}\right)\left|\mu,\,\tau\right> \label{eq:cotrspec} \\
& = & \frac{\sqrt{\gamma}}{2} l_{p} |\mu| \left( \sqrt{|\tau+1|} - \sqrt{|\tau-1|}\right)\left|\mu,\,\tau\right> \,, \nonumber \\
\hat{\omega}_{2} \left|\mu,\,\tau\right> &=& \sqrt{\gamma} l_{p} \mbox{sgn}(\mu) \sqrt{|\tau|} \left|\mu,\,\tau\right> . \nonumber
\end{eqnarray}
 Note that the $\hat{\omega}_{2}$ eigenvalues have the same relation to $E^{R}$'s eigenvalues as in the classical relationship. However, $\hat{\omega}^{R}$'s eigenvalues relation to the triad's eigenvalues deviates from classical relation and only asymptotically approaches the classical relation for large $\tau$. Since the classical singularity is located at $\tau=0$, quantum effects become important near the singularity. It may be seen from the above that the co-triad is \emph{not} singular in the quantum theory  and $\tau=0$ is well behaved.

The evolution of the Hamiltonian constraint has also been used to argue the elimination of the classical singularity \cite{ref:AandB}. Utilising the quantum operators, the Hamiltonian constraint can be constructed. The Hamiltonian constraint operator, acting on the states yields a \emph{difference} equation for $C_{\mu\tau}$ of the form \cite{ref:AandB}:
\begin{align}
&\hat{H}^{\delta}_{\mbox{\tiny{grav}}}\,C_{\mbox{\tiny{$\mu,\tau$}}}=2\delta \left(\sqrt{|\tau+2\delta|} +\sqrt{|\tau|}\right) \left[C_{\mbox{\tiny{$\mu+2\delta,\tau +2\delta$}}} \right. \nonumber \\
& \left. - C_{\mbox{\tiny{$\mu-2\delta,\tau+2\delta$}}}\right]
 +\left(\sqrt{|\tau+\delta|} - \sqrt{|\tau-\delta|}\right) \left[(\mu+2\delta) C_{\mbox{\tiny{$\mu+4\delta,\tau$}}}  \right. \nonumber \\ 
&\left. -(1+2\gamma^{2}\delta^{2})\mu\, C_{\mbox{\tiny{$\mu,\tau$}}} + (\mu -2\delta)C_{\mu-4\delta, \tau} \right]  \nonumber \\
&+2\delta\left(\sqrt{|\tau-2\delta|} +\sqrt{|\tau|}\right) \left[C_{\mbox{\tiny{$\mu-2\delta,\tau-2\delta$}}} 
- C_{\mbox{\tiny{$\mu+2\delta,\tau-2\delta$}}}\right]=0 \:. \label{eq:hdif}
\end{align} 
The quantity $\delta$ is sometimes known as the holonomy parameter, which measures the size of the steps taken in the $R$ and angular directions ($\delta$ is assumed to be the same in all directions here.)

 The particular expression (\ref{eq:hdif}) is valid for $\mu \geq 4\delta$ but similar expressions exist for other values of $\mu$. The form of the constraint for $\mu < 4$ dictates that only data at $\mu=0$ and $\mu=\delta$ needs to be specified and the issue of singularity resolution turns out to be insensitive to the choice of initial data \cite{ref:AandB}. 

The evolution  described by (\ref{eq:hdif}) procedes smoothly as long as the coefficient of the wavefunction is non-zero at $\tau-2\delta$. The coefficient here is $\sqrt{|\tau-2\delta|}+\sqrt{|\tau|}$ and never vanishes. The evolution is therefore smooth across the singularity. It turns out that the analogous Wheeler-deWitt equation cannot be evolved beyond the point corresponding to the classical singularity.

Before proceeding it is worth noting that some issues exist which make the above analysis potentially suspect. It has been argued that by considering $\delta$ to be a constant the above system does not possess an good semi-classical limit \cite{ref:5inBV}. In the field of loop quantum cosmology, where similar studies are performed, the parameter analogous to $\delta$ must depend explicitly on the scale factor to get an appropriate semi-classical limit. Therefore, in \cite{ref:BandV}, the condition of constant $\delta$ was relaxed. The approach also relied on an effective Hamiltonian approach, whose equations of motion capture the essence of the main quantum corrections. This is more in-line with approaches taken in loop quantum cosmology and was first applied to black holes in \cite{ref:11inBandV} for the case of constant $\delta$. For constant $\delta$, the effective dynamics are encoded in the modified Hamiltonian (compare with (\ref{eq:frozenham})):
\begin{align}
 \mathcal{H}_{eff}=&-\frac{N}{2\gamma^{2}}\left[2\frac{\sin(A_{2}\delta)\,\,\sin(A_{R}\delta) }{\delta^{2}}\sqrt{E^{R}} \right. \nonumber \\
 &\left. +\left( \frac{\sin^{2} A_{2}\delta}{\delta^{2}} + \gamma^{2}\right) \frac{E^{2}}{\sqrt{E^{R}}}\right]\,. \label{eq:effham1}
\end{align}
An appropriate choice of lapse function, $N$, allows one to solve the resulting system of Hamilton equations for $A_{R},\, A_{2},\, E^{R}$ and $E^{2}$. The mathematical result may be found in \cite{ref:11inBandV} and \cite{ref:BandV}. Qualitatively, the solutions describe a bounce in place of the classical singularity. This bounce connects two (non-singular) black holes of generally unequal mass.

The above results, though promising, depend on a fixed holonomy parameter, which has been shown to be problematic in the semi-classical limit when similar fixed $\delta$ techniques were used in loop quantum cosmology. Therefore an improved scheme was sought in \cite{ref:BandV}. Following loop quantum cosmology, non fixed $\delta$ quantities are utilized instead. The natural question that arises then is how to set the fundamental size of the two $\delta$ parameters that appear in the improved theory. We denote the holonomy parameter associated with the ``radial'' direction as $\delta_{R}$ and the holonomy parameter associated with the angular direction as $\delta_{2}$, to make ties with the notation above.

The holonomies in the Hamiltonian involve closed loops and one may constrain the size of the loop by appealing to the fact that the holonomy loops make up the minimum area allowed by loop quantum gravity \footnote{There are various schemes to impose a non-fixed $\delta$ and we only concentrate on one here for brevity.}:
\begin{equation}
 Ar_{0}=4\sqrt{3}\pi l_{p}^{2}\gamma\,.
\end{equation}
The area of a $\theta - \phi$ surface element is given by
\begin{equation}
 Ar_{\theta\phi}=E^{R} \left(\delta_{2}\right)^{2}\,,
\end{equation}
which, by restricting that this equal the minimum area, yields
\begin{equation}
 \delta_{2}=\sqrt{\frac{Ar_{0}}{E^{R}}}\,.
\end{equation}
Similarly, given that the area of an $R-\theta$ surface element is given by $Ar_{R\theta}=E^{2}\delta_{R}\delta_{2}$, one arrives at the condition
\begin{equation}
 \delta_{R}=\frac{\sqrt{Ar_{0}\,E^{R}}}{E^{2}}\,.
\end{equation}
The generalization of (\ref{eq:effham1}) to form the Hamiltonian constraint is straight-forward:
\begin{align}
 \mathcal{H}_{eff}=&-\frac{N}{2\gamma^{2}}\left[2\frac{\sin(A_{2}\delta_{2})\,\,\sin(A_{R}\delta_R) }{\delta_{2}\delta_{R}}\sqrt{E^{R}} \right. \nonumber \\
&\left.  +\left( \frac{\sin^{2} A_{2}\delta_{2}}{(\delta_{2})^{2}} + \gamma^{2}\right) \frac{E^{2}}{\sqrt{E^{R}}}\right]\,. \label{eq:effham2}
\end{align}
The resulting equations of motion were solved numerically in \cite{ref:BandV}. There, the authors started the evolution near the horizon, using initial data compatible with the classical solution, along with the Hamiltonian constraint $\mathcal{H}=0$. As in the previous scenarios, the singularity is avoided, with 2-sphere radii being bounded from below. In this improved scheme the solution does not connect two black holes but instead asymptotes to a Nariai solution at late time. For reference, the type of Nariai solution compatible with this evolution takes the form
\begin{equation}
 ds^{2}=a^{2}\left(d\tau^{2} +\cosh^{2}\tau\,\,d\chi^{2}\right) +b^{2}\,d\Omega^{2}\,,
\end{equation}
with $a$ and $b$ constants. Several figures are presented below in figure \ref{fig:stevesfigs1} to show a sample evolution (see figure caption for details).

\begin{figure}[ht!]
\begin{center}
\includegraphics[bb=10 122 646 696, clip, scale=0.30, keepaspectratio=true]{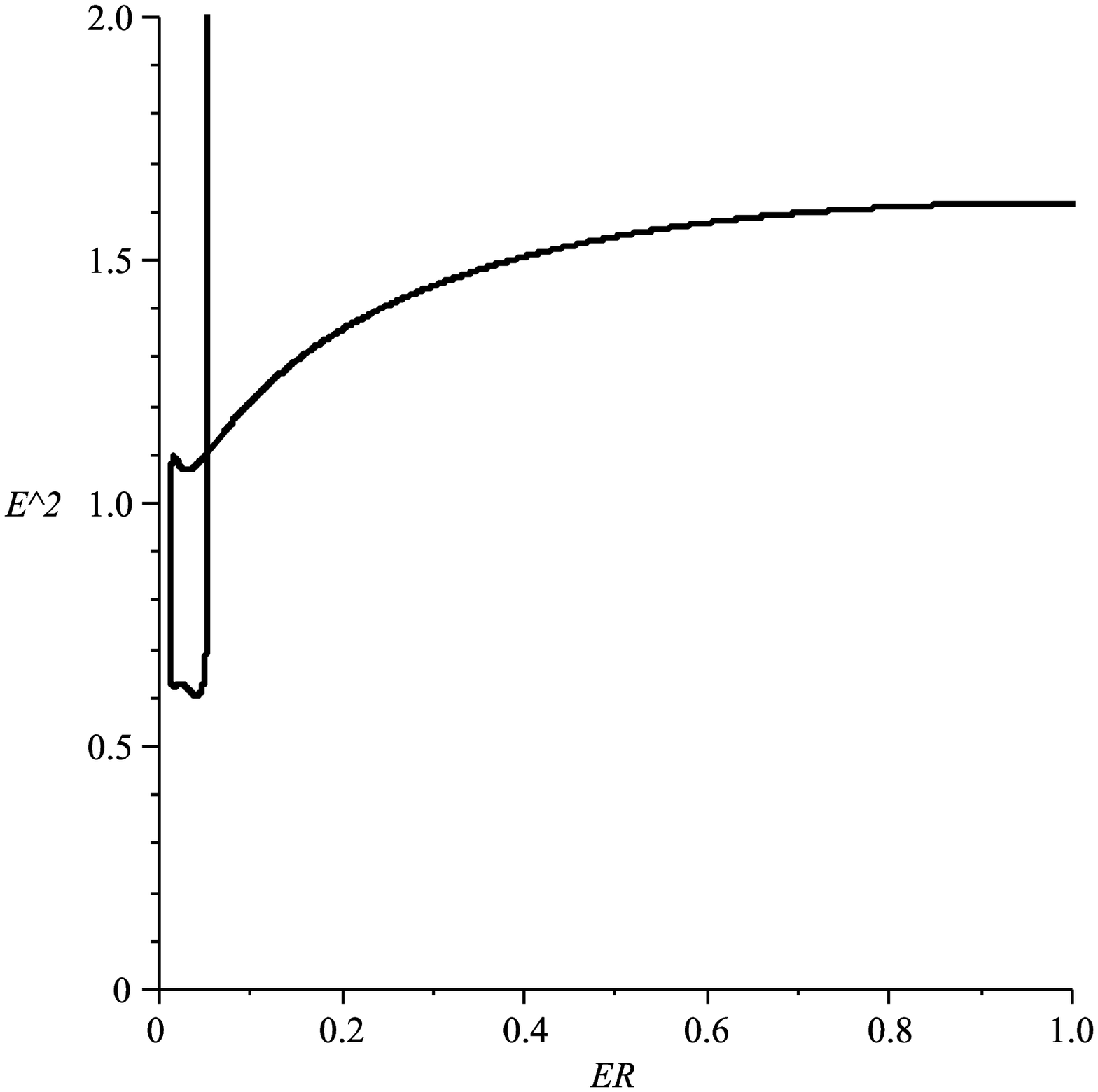} 
\includegraphics[bb=18 102 624 682, clip, scale=0.30, keepaspectratio=true]{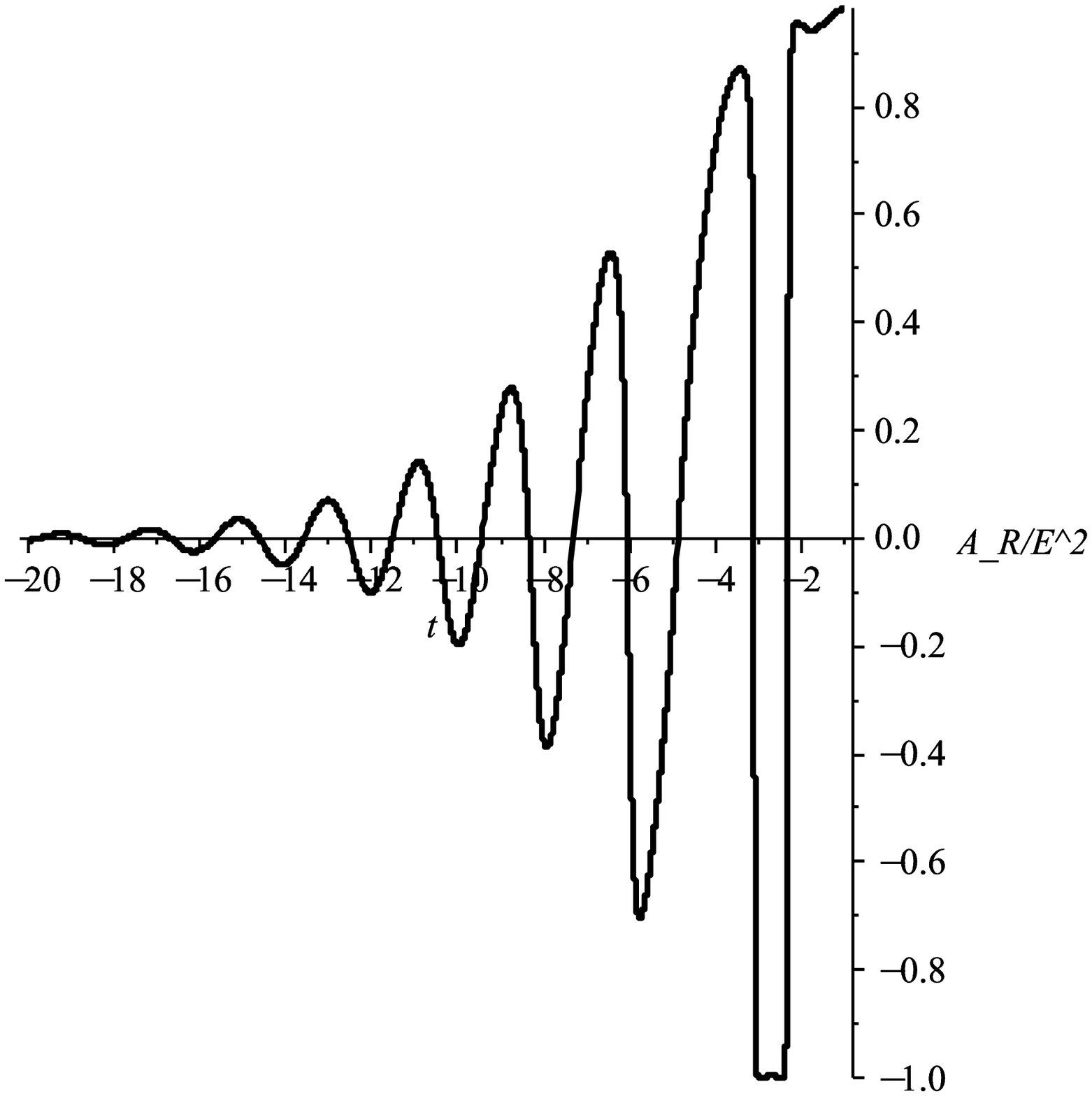}
\caption{{\small A sample evolution utilizing the improved quantization technique in \cite{ref:BandV}. The top plot relates the two densitized triad variables, $E^{R}$ and $E^2$. At the classical singularity, the curves would intersect the origin. This does not happen in the quantum evolution. On the bottom, the combination $A^{R}/E^{2}$ as a function of the Hamiltonian evolution time $t$ ($T=e^{t}$). This remains finite and approaches a constant for $t\rightarrow -\infty$. This quantity would become unbounded at a curvature singularity. I thank Steve Kloster for generating the figures.}} \label{fig:stevesfigs1}
\end{center}
\end{figure}

\section{{\small Concluding remarks}}
We have briefly summarized here the status of black hole singularity studies within loop quantum gravity. It should be noted that at this stage it is not clear if the limitations of the approximations above are serious enough to disqualify all of the results. A discussion of the drawbacks and strengths to various methods within the framework of loop quantum cosmology may be found in the recent paper \cite{ref:paperofsteve}. However, it is encouraging that all the methods indicate that the classical singularity is indeed avoided in the evolution. One may say with some confidence that the singularity may therefore be avoided within the full theory.

Future studies of interest would be to study other systems which are singular in classical gravity as well as the incorporation of higher order effects. Of particular interest in all of these studies is what exactly replaces the classical singularity. This is certainly one of the most engaging questions a quantum gravity theory can answer.

In closing I would like to thank Richard MacKenzie, Arundhati Dasgupta, and all the other personnel involved with the organization of the Theory Canada IV conference.

\begin{spacing}{0.2}
 \linespread{0.2}
\bibliographystyle{unsrt}

\end{spacing}

\end{document}